1# Segmentation of Muscularis Propria in Colon Histopathology Images Using Vision Transformers for Hirschsprung's Disease

Youssef Megahed[1], Anthony Fuller[1], Saleh Abou-Alwan[1], Dina El Demellawy[2], Adrian D. C. Chan[1]

[1]Department of Systems and Computer Engineering, Carleton University, Ottawa, Canada
[2]Children's Hospital of Eastern Ontario (CHEO), Ottawa, Canada

*Abstract*—**Hirschsprung's disease (HD) is a congenital birth defect diagnosed by identifying the lack of ganglion cells within the colon's muscularis propria, specifically within the myenteric plexus regions. There may be advantages for quantitative assessments of histopathology images of the colon, such as counting the ganglion and assessing their spatial distribution; however, this would be time-intensive for pathologists, costly, and subject to inter- and intra-rater variability. Previous research has demonstrated the potential for deep learning approaches to automate histopathology image analysis, including segmentation of the muscularis propria using convolutional neural networks (CNNs). Recently, Vision Transformers (ViTs) have emerged as a powerful deep learning approach due to their self-attention. This study explores the application of ViTs for muscularis propria segmentation in calretinin-stained histopathology images and compares their performance to CNNs and shallow learning methods. The ViT model achieved a DICE score of 89.9% and Plexus Inclusion Rate (PIR) of 100%, surpassing the CNN (DICE score of 89.2%, PIR of 96.0%) and k-means clustering method (DICE score of 80.7%; PIR 77.4%). Results assert that ViTs are a promising tool for advancing HD-related image analysis.**

## INTRODUCTION

Hirschsprung disease (HD) is a congenital defect involving the absence of ganglion cells that should be found within myenteric plexus regions [15], within muscularis propria section of the colon (Figure 1). HD has a prevalence rate of 1 in 5000 infants and can be fatal if untreated [14]. The *pull-through procedure is* a surgical intervention that removes the aganglionic section of the colon and joins the healthy portion to the anus.

Discerning healthy and aganglionic sections of the colon is performed by pathologists who visually assess histopathology images of the colon. Quantitative analyses, such as counting ganglion cells and assessment of their spatial distribution, may lead to increased surgical success and better patient outcomes by establishing clear criteria for discerning healthy and aganglionic sections. However, such quantitative assessment would be time-consuming for pathologists and would increase healthcare costs. In addition, manual assessment is prone to inter- and intra-rate variability [1, 2].

Attempts have been made to automate the analysis of histopathology whole slide images (WSIs) of the colon, where the detection of ganglion cells was divided into three-stages [3-6]: (1) segmenting the muscularis, (2) within the muscularis, segmenting the plexus regions, and (3) within the plexus regions, identifying ganglion cells. In [6], k-means clustering, a shallow machine learning approach, was used for segmenting the muscularis, achieving a DICE score of 70.7% and a mean Plexus Inclusion Rate (PIR) for myenteric plexus of 77.4% (percentage of all plexus regions that are found within the segmented muscularis region). Recently, deep learning approaches have demonstrated strong potential in medical image analysis, including computation pathology [13]. A Convolutional Neural Network (CNN)-based model, a popular deep-learning approach, was used for muscularis segmentation [5, 6], achieving a DICE coefficient of 89.2% and a mean PIR of 96.0%.

Although CNNs are a predominant model architecture within computer vision, Vision Transformers (ViTs) are gaining popularity, often outperforming CNNs. CNNs focus on local relationships via convolutions, whereas ViTs can capture long-range relationships using the self-attention mechanism. The objective of this study is to examine ViTs for segmenting the muscularis region, which is anticipated to outperform CNN-based models.

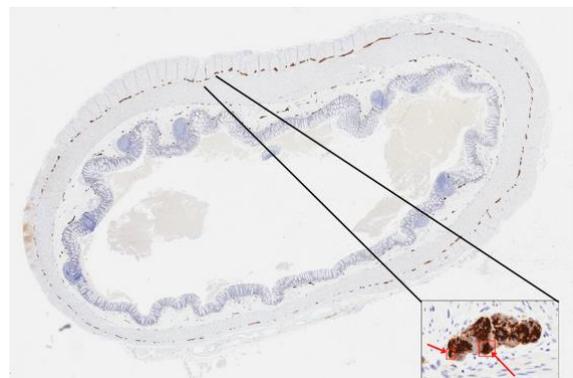

Figure 1: Whole slide image of a cross-section of the colon. Zoomed-in portion shows a plexus region with ganglion cells indicated with red arrows.

## METHODOLOGY

### Data

The dataset (WSIs and ground truth annotations) used in this study is the same dataset used in [6]. Data consists of 30 WSIs from 26 patients diagnosed with HD. Images were acquired by the Children's Hospital of Eastern Ontario (CHEO) using the digital scanner *Aperio Scan Scope CS* at 20× resolution (0.50 μm/pixel). Each WSI is associated with three ground truth annotations: (i) the muscularis propria, which was manually segmented, (ii) the myenteric plexus regions, which were roughly manually segmented (i.e., a visually noticeable amount of tissue around the plexus regions are also included), and (iii) the ganglion cells, which were roughly manually segmented (i.e., parts of the ganglion cell may be excluded and/or areas around the cell may be included, as it can be difficult to visually identify the exact border of a ganglion cell). A confidence level accompanies each ganglion cell annotation. A high confidence level indicates a high certainty that the annotated object is a ganglion cell. A low confidence level indicates a belief that the annotated object is a ganglion cell, but there is some uncertainty.

### Pre-processing

Color variance among the WSIs, likely due to differences in the staining procedures, was mitigated via the Macenko colour normalization [8]. WSIs were also downsampled by a factor of 10 (resultant resolution of 5 μm/pixel).

### Deep Learning Model

A ViT model, pre-trained on the ImageNet-1k dataset via masked autoencoding [12], and then fine-tuned (see Model Training and Testing section), was employed. The ViT processes input tiles (size 224×224) by dividing them into non-overlapping 16×16 pixel patches [7]. Each patch is transformed into a feature representation of size 768 and fed through multiple transformer encoder layers that use self-attention and feed-forward networks. These mechanisms progressively encode morphological details essential for distinguishing structures, such as the muscularis propria and plexus regions. The output layer comprised of a fully connected linear segmentation head, mapping each 16×16 patch embedding to binary pixel-wise predictions (i.e., muscularis or not muscularis). The binary segmentation maps were generated by applying a SoftMax function across the logits and using a confidence threshold value to classify each pixel (the confidence threshold values were investigated in Figure 2).

### Model Training and Testing

We used a 5-fold cross-validation, splitting the 30 WSIs into groups of 6. Within each fold, 5 groups (24 WSIs) were used to train the model, and 1 group (6 WSIs) was used to test the model (i.e., segment the muscularis). The test group was rotated in each fold, so all WSIs were eventually tested.

For training, WSIs were tiled into overlapping smaller sub-images of size 224×224 pixels. From each WSI in the training set, 1000 tiles were randomly sampled, resulting in a total of 24,000 tiles. Data augmentation techniques were applied during training, including random rotations, horizontal and vertical flips, and scaling. The model was optimized using AdamW with a base learning rate of 5e-4 (the initial learning rate before adjustments) and a weight decay of 1e-4. A cosine learning rate schedule, which adjusts the learning rate over time, with a warmup period of five epochs, was employed. The model was trained for 50 epochs with a batch size of 64.

For testing, segmentation was performed using overlapping tiles across each WSI (224×224 pixels tile size and 112 pixel stride). After each tile was segmented by the ViT, the center segmentation of each tile (size 112×112 pixels) was stitched together to form the final WSI segmentation map.

### Performance Metrics

The DICE score is used to evaluate segmentation accuracy, which assesses the ratio of overlapping regions:

$$DICE = \frac{2|X_{ViT} \cap Y_{GT}|}{|X_{ViT}| + |Y_{GT}|} \quad (1)$$

where $X_{ViT}$ and $Y_{GT}$ are the set of muscularis pixels in the binary masks of ViT and ground truth manual segmentations, respectively. A DICE score of 100% signifies complete overlap between the predicted and ground truth masks, whereas a value of 0% denotes no overlap.

Table 1: Performance comparison (%) of segmentation models on muscularis propria.

| Model | DICE | Precision | Recall | PIR |
|---|---|---|---|---|
| k-means [6] | 70.7 | 70.6 | 78.9 | 77.4 |
| CNN [5,6] | 89.2 | 81.9 | 96.2 | 96.0 |
| **ViT (ours)** | **89.9** | **82.4** | **99.7** | **100** |

Segmentation accuracy was also evaluated using the Plexus Inclusion Rate (PIR):

$$PIR = \frac{N_{ViT}}{N_{GT}} \quad (2)$$

where $N_{ViT}$ and $N_{GT}$ are the number of plexus regions in the muscularis of the ViT and ground truth manual segmentations, respectively. As plexus regions are found toward the middle of the muscularis, segmentation errors toward the outer regions of the muscularis may not be important, as the ultimate goal is to identify ganglion cells within the plexus regions. As such, it is important for the muscularis segmentation to not exclude plexus regions.

$$Precision = \frac{TP}{TP + FP} \quad (3)$$

$$Recall = \frac{TP}{TP + FN} \quad (4)$$

Precision and recall pixel-wise metrics were also used to evaluate segmentation accuracy, where positive is associated with a pixel being part of the muscularis (e.g., a false negative FN indicates the ViT classified a pixel as not being muscularis when the ground truth annotation did).

## RESULTS

Segmentation results (Table 1) demonstrate the superiority of the ViT model, compared to the k-means [6] and CNN models [5,6], across all metrics (i.e., DICE, precision, recall, and PIR). The ViT model achieved a PIR of 100%, which was much higher than the k-means (77.4%) and CNN (96.0%) models. Figure 2 shows the impact of the confidence threshold on the DICE score and PIR. As the confidence threshold decreases, the ViT segments more pixels as part of muscularis. At the 0.4 threshold, the DICE reaches a maximum of 95.4%. Decreasing the threshold further increases the PIR reaching 100% at a threshold of 0.01; the associated DICE, precision, and recall are 89.9%, 82.4%, and 99.7%, respectively.

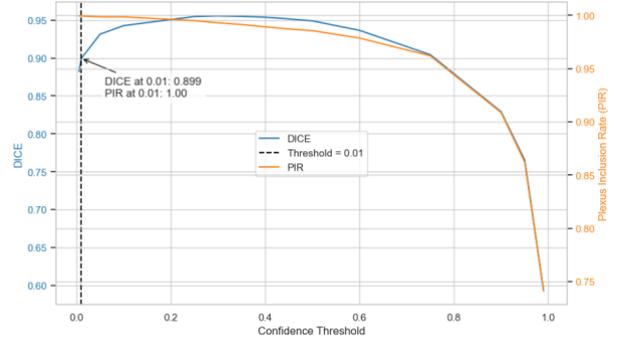

Figure 2: Impact of confidence threshold on DICE score and plexus inclusion rate.

## DISCUSSION

Results demonstrate that ViTs provide a promising approach for segmenting the muscularis propria in calretinin-stained histopathology images, a task critical in diagnosing Hirschsprung's disease. ViTs achieve high segmentation accuracy and PIR compared to CNN and shallow learning models. A PIR of 100% is obtainable with a relatively high DICE score (89.9%).

While a higher DICE score is possible, the decrease in PIR is a poor tradeoff. Perfect PIR means the muscularis segmentation did not exclude any plexus regions; exclusion of a plexus region would result in ganglion cells within that plexus region not being detected in later processing stages. The lower DICE score associated with the 100% PIR is associated with a lower precision; the ViT identified more pixels as being muscularis that were not muscularis in the ground truth annotations. The implication of this lower segmentation accuracy is that the next processing stage (plexus segmentation) would have to search across a larger area (i.e., the segmented muscularis); however, the segmentation is quite accurate when the PIR is 100%, so the computation costs associated with that increased search area is small.

## CONCLUSION

This study presented a ViT-based approach for segmenting the muscularis propria in calretinin-stained histopathology images relevant to Hirschsprung's disease diagnosis. Our method outperformed traditional shallow learning k-means and deep-learning CNN-based models, demonstrating high segmentation accuracy and PIR. These results underscore the potential of ViTs as a robust and precise tool in histopathological segmentation tasks. The success of our ViT model in muscularis segmentation suggests that it could be a promising approach for the remaining stages of automated Hirschsprung's disease analysis; specifically, plexus segmentation and ganglion detection.

If accurate ganglion detection can be achieved, it could potentially empower diagnostic workflows with increased reliance on automated pathology by appropriately identifying and segmenting these critical anatomical regions. It would enable clinical research that can correlate ganglion cell features (e.g., count and spatial distribution) to surgical outcomes, which in turn could improve the efficiency and efficacy of interventions for Hirschsprung's disease.

## REFERENCES

[1] A. Mukherjee *et al.*, "The placental distal villous hypoplasia pattern: interobserver agreement and automated fractal dimension as an objective metric," *Pediatric and Developmental Pathology*, vol. 19, no. 1, pp. 31–36, 2016.

[2] A. J. Demetris *et al.*, "Intraobserver and interobserver variation in the histopathological assessment of liver allograft rejection," *Hepatology*, vol. 14, no. 5, pp. 751–755, 1991.

[3] J. Kurian *et al.*, "Image Processing and Analysis of Histopathological Images Relating to Hirschsprung's Disease," *CMBES Proceedings*, vol. 41, 2018.

[4] M. T. K. Law, A. D. C. Chan, and D. El Demellawy, "Color image processing in Hirschsprung's disease diagnosis," in *2016 IEEE EMBS International Student Conference (ISC)*, 2016, pp. 1–4.

[5] C. McKeen, F. Zabihollahy, J. Kurian, A. D. Chan, D. El Demellawy, and E. Ukwatta, "Machine learning-based approach for fully automated segmentation of muscularis propria from histopathology images of intestinal specimens," in Medical Imaging 2019: Digital Pathology, vol. 10956, pp. 146–151, Mar. 2019.

[6] J. A. Kurian, Automated Identification of Myenteric Ganglia in Histopathology Images for the Study of Hirschsprung's Disease, M.A.Sc. thesis, Carleton University, 2021.

[7] M. Minderer, G. Heigold, S. Gelly, J. Uszkoreit, and N. Houlsby, "An Image is Worth 16x16 Words: Transformers for Image Recognition at Scale," arXiv preprint arXiv:2010.11929, 2020. [Online]. Available: https://arxiv.org/abs/2010.11929.

[8] M. Macenko *et al.*, "A method for normalizing histology slides for quantitative analysis," in *2009 IEEE International Symposium on Biomedical Imaging: From Nano to Macro*, 2009, pp. 1107–1110.

[9] A. Atabansi, et al., "Applications of Transformers in Histopathological Image Analysis: A Comprehensive Survey," *Biomedical Engineering Online*, vol. 23, no. 1, 2023. DOI: 10.1186/s12938-023-01157-0.

[10] A. Kanadath, J. A. A. Jothi, and S. Urolagin, "CViTS-Net: A CNN-ViT Network with Skip Connections for Histopathology Image Classification," *IEEE Access*, 2024.

[11] L. Hörst, et al., "CellViT: Vision Transformers for Precise Cell Segmentation and Classification," *arXiv preprint*, 2023. [Online]. Available: https://arxiv.org/abs/2306.15350.

[12] K. He et al., "Masked Autoencoders Are Scalable Vision Learners," *Proceedings of the IEEE/CVF Conference on Computer Vision and Pattern Recognition (CVPR)*, 2022, pp. 16000–16009. doi: 10.1109/CVPR52688.2022.01554.

[13] J. van der Laak, G. Litjens, and F. Ciompi, "Deep learning in histopathology: the path to the clinic," *Nature Medicine*, vol. 27, no. 5, pp. 775–784, 2021.

[14] J. Kessmann, "Hirschsprung disease: diagnosis and management," American Family Physician, vol. 74, no. 8, pp. 1319-1322, 2006.

[15] S. Lotfollahzadeh, M. Taherian, and S. Anand, "Hirschsprung disease," in *StatPearls*, StatPearls Publishing, 2023.